# Universal machine learning interatomic potentials poised to supplant DFT in modeling general defects in metals and random alloys

Fei Shuang[a*], Zixiong Wei[a], Kai Liu[a], Wei Gao[b,c], Poulumi Dey[a*]

[a]Department of Materials Science and Engineering, Faculty of Mechanical Engineering, Delft University of Technology, Mekelweg 2, 2628 CD Delft, The Netherlands

[b]J. Mike Walker'66 Department of Mechanical Engineering, Texas A&M University, College Station, TX 77843, United States

[c]Department of Materials Science & Engineering, Texas A&M University, College Station, TX 77843, United States

∗ Corresponding authors.

E-mail address: P.dey@tudelft.nl (P. Dey), F.Shuang@tudelft.nl (F. Shuang)

**Abstract:** Recent advances in machine learning, combined with the generation of extensive density functional theory (DFT) datasets, have enabled the development of universal machine learning interatomic potentials (uMLIPs). These models offer broad applicability across the periodic table, achieving first-principles accuracy at a fraction of the computational cost of traditional DFT calculations. In this study, we demonstrate that state-of-the-art pretrained uMLIPs can effectively replace DFT for accurately modeling complex defects in a wide range of metals and alloys. Our investigation spans diverse scenarios, including grain boundaries and general defects in pure metals, defects in high-entropy alloys, hydrogen-alloy interactions, and solute-defect interactions. Remarkably, the latest EquiformerV2 models achieve DFT-level accuracy on comprehensive defect datasets, with root mean square errors (RMSE) below 5 meV/atom for energies and 100 meV/Å for forces, outperforming specialized machine learning potentials such as moment tensor potential and atomic cluster expansion. We also present a systematic analysis of accuracy versus computational cost and explore uncertainty quantification for uMLIPs. A detailed case study of tungsten (W) demonstrates that data on pure W alone is insufficient for modeling complex defects in uMLIPs, underscoring the critical importance of advanced machine learning architectures and diverse datasets, which include over 100 million structures spanning all elements.

These findings establish uMLIPs as a robust alternative to DFT and a transformative tool for accelerating the discovery and design of high-performance materials.

## 1 Introduction

Machine learning is revolutionizing computational materials science across multiple dimensions, notably enhancing predictive modeling capabilities and accelerating materials discovery [1–8]. One of the most significant achievements in this transformative era is the development of universal machine learning potentials (uMLIPs). These potentials represent a paradigm shift in how scientists conduct first-principle accurate simulations across the periodic table. Unlike specialized MLIPs (sMLIPs) that require extensive recalibration for each new element or compound and substantial training, uMLIPs offer ready-to-use models that deliver unprecedented accuracy and transferability in predicting energy, force, and stress. Over the past years, uMLIPs have rapidly evolved. As of the execution of this work, the Matbench-discovery repository catalogues 17 distinct uMLIPs, each featuring its unique set of models and varying parameter counts [9]. Notable architectures include Voronoi RF [10], BOWSR [11], Wrenformer [12], CGCNN+P [13], CGCNN [14], MEGNet [15], ALIGNN [16], M3GNet [17], CHGNet [18], MACE [19], GRACE [20], SevenNet [21], Orb [22], GNoME [23], MatterSim [24], EquiformerV2 (eqV2) [25,26], and DPA3 [27,28].

The primary objective of the uMLIPs is to supplant computationally expensive DFT calculations, a goal that remains substantially unachieved, since the uncertainty regarding the accuracy of uMLIPs presents a major hurdle. To address this, systematic benchmark studies have been conducted by researchers in various fields. One of the most comprehensive of these was performed by Deng et al. [29], who observed a consistent softening phenomenon across a range of scenarios including surfaces, defects, solid-solution energetics, phonon vibration modes, ion migration barriers, and high-energy states in M3GNet, CHGNet, and MACE-MP-0. This softening originates primarily from the systematically underpredicted potential energy surface curvature, which was attributed to the biased sampling of near-equilibrium configurations in the uMLIPs' pre-training datasets. Another study assessed the performance of CHGNet, M3GNet, MACE-MP-0, and ALIGNN in terms of the equation of state, structural optimization, formation energy, and vibrational properties, noting variable convergence rates during relaxation among the different uMLIPs [30]. Further research focused on the predictive capabilities of MACE-MP-0, CHGNet, and M3GNet regarding surface energy [31], as well as the accuracy of these models in determining mixing enthalpies and volumes of disordered alloys and complex high-entropy alloys. These findings collectively highlight considerable uncertainties in the performance of existing uMLIPs, particularly concerning defects in solid materials. Recommendations from these studies suggest that fine-tuning uMLIPs by including relevant data can enhance their accuracy. However, the amount of data required for this fine-tuning remains unclear. If an extensive amount of data is necessary, comparable to that needed for training sMLIPs, uMLIPs may not offer a competitive advantage. Furthermore, fine-tuning uMLIPs demands domain expertise in DFT calculations and access to high-performance computational resources, especially Graphics Processing Units (GPUs), which may not be accessible to all researchers.

Fortunately, the recent release of four advanced uMLIPs, including Orb, MatterSim, eqV2, and DPA3 represents a significant advancement in the Matbench-discovery repository [9]. These models, which involve tens of millions of parameters and utilize training datasets comprising hundreds of millions of structures, establish a new benchmark for accuracy. In this study, we leverage these innovations to evaluate the reliability of uMLIPs in modeling complex defects in metals and alloys. To this end, we have generated and collected DFT datasets featuring extensive defects in five pure metals (Mo, Nb, Ta, W, and Mg) and a range of alloys from low-to-high entropy (MoNb, CrCoNi, MoNbTaW, HEA10-AlHfMoNbNiTaTiVWZr), as well as metals and alloys containing interstitial atoms (MoNbTaW-H). These datasets enable a comprehensive assessment of the fidelity with which uMLIPs model common defects in metals and alloys. A systematic analysis of accuracy versus computational cost and uncertainty quantification analysis are provided. Additionally, we explore the differences between uMLIPs and sMLIPs in terms of machine learning architecture and training datasets. This analysis helps to elucidate the unexpectedly high accuracy of uMLIPs in modeling defects in these materials.

## 2 Methods

We utilize the Vienna Ab initio Simulation Package (VASP) to perform first-principles calculations of all new configurations [32]. A gradient-corrected functional in the Perdew-Burke-Ernzerhof (PBE) form is used to describe the exchange and correlation interactions [33]. Electron-ion interactions are treated within the projector-augmented-wave (PAW) method, using the standard PAW pseudopotentials provided by VASP [34]. The energy convergence criterion is set to $10^{-6}$ eV for electronic self-consistency calculations. The plane-wave cutoff energy is chosen to be 520 eV. The KPOINTS are generated by VASPKIT [35], based on the Monkhorst-Pack scheme, with a consistent density of $2\pi\times0.03$ $Å^{-1}$. Collinear spin-polarization is included for all magnetic species. These computational parameters are applied consistently to every dataset produced in this work. The Atomic Simulation Environment (ASE) was used for all one-shot calculations involving uMLIPs [36]. Additionally, OVITO was employed for the visualization of the atomic structures [37].

In this study, we evaluate the performance of seven of state-of-the-art uMLIPs, encompassing a total of 26 models as detailed in Table 1. Given the proven accuracy of prior uMLIPs such as M3GNet and SevenNet, our analysis focuses exclusively on models that have demonstrated robust performance in Table 1. CHGNet is distinguished by its unique ability to predict magnetic moments. MACE is represented by three models: MACE-MP-0, MACE-MPA-0, and MACE-omat-0, where “MP” corresponds to the MPtrj dataset [18], “MPA” combines the MPtrj dataset with the subsampled Alexandria dataset [38], and “omat” represents the open materials dataset 2024 for inorganic materials (OMat24) [25]. MatterSim includes two models, MatterSim-v1.0.0-1M and MatterSim-v1.0.0-5M, each differing in parameter size and trained on newly developed DFT datasets that span temperatures from 0 to 5000 K and pressures up to 1000 GPa. Orb features four models, each with distinct training datasets and complexities. eqV2, the most extensive group, consists of ten models employing comprehensive DFT datasets including MPtrj, the subsampled Alexandria (sAlex), and OMat24. It is noteworthy that MPtrj and the subsampled Alexandria datasets contain near-equilibrium configurations, whereas the OMat24 datasets consist of non-equilibrium configurations [25]. The Denoising Non-equilibrium Structure (DeNS) protocol was utilized to enhance the performance of eqV2 models (eqV2-dens models) [39]. GRACE is represented by four models: GRACE-1L-r6-MP and GRACE-2L-r6, which are trained on the MPtrj dataset, and GRACE-1L-OAM and GRACE-2L-OAM, which are trained on three datasets: OMat24, MPtrj, and sAlex. Lastly, DPA3-v1 is a large-scale atomic model implemented in the DeePMD-kit package [28]. Two recently released models, DPA3-v1-MPtrj and DPA3-v1-OpenLAM, have demonstrated outstanding performance in the Matbench-discovery repository [9]. In the following section, we will assess the performance of these 26 models.

Table 1 Different uMLIPs and corresponding models, training datasets, structures and parameters.

| ID | Model name | Training dataset | No. of structures | No. of parameters |
|---|---|---|---|---|
| 1 | CHGNet_0.3.0 | MPtrj | 1.58M | 413k |
| 2 | MACE-MP-0 (large) | MPtrj | 1.58M | unknown |
| 3 | MACE-MPA-0 | MPtrj+sAlex | 12M | 9.06M |
| 4 | MACE-omat-0 | OMat24 | 101M | 9.06M |
| 5 | MatterSim-v1.0.0-1M | MatterSim | 17M | 1M |
| 6 | MatterSim-v1.0.0-5M | MatterSim | 17M | 4.55M |
| 7 | Orb-MPtraj-only-v2 | MPtrj | 1.58M | 25.2M |
| 8 | Orb-d3-xs-v2 (5 layers) | MPtrj+Alex | 32.1M | unknown |
| 9 | Orb-d3-sm-v2 (10 layers) | MPtrj+Alex | 32.1M | unknown |
| 10 | Orb-d3-v2 | MPtrj+Alex | 32.1M | 25.2M |
| 11 | eqV2-31M-mp | MPtrj | 1.58M | 31M |
| 12 | eqV2-dens-31M-mp | MPtrj | 1.58M | 31M |
| 13 | eqV2-dens-86M-mp | MPtrj | 1.58M | 86M |
| 14 | eqV2-dens-153M-mp | MPtrj | 1.58M | 153M |
| 15 | eqV2-31M-omat | OMat24 | 101M | 31M |
| 16 | eqV2-86M-omat | OMat24 | 102M | 86M |
| 17 | eqV2-153M-omat | OMat24 | 102M | 153M |
| 18 | eqV2-31M-omat-mp-salex | OMat24+MPtrj+sAlex | 113M | 31M |
| 19 | eqV2-86M-omat-mp-salex | OMat24+MPtrj+sAlex | 113M | 86M |
| 20 | eqV2-153M-omat-mp-salex | OMat24+MPtrj+sAlex | 113M | 153M |
| 21 | GRACE-1L-r6-MP | MPtrj | 1.58M | unknown |
| 22 | GRACE-2L-r6-MP | MPtrj | 1.58M | 15.3M |
| 23 | GRACE-1L-OAM | OMat24+MPtrj+sAlex | 113M | 3.45M |
| 24 | GRACE-2L-OAM | OMat24+MPtrj+sAlex | 113M | 12.6M |
| 25 | DPA3-v1-MPtrj | MPtrj | 1.58M | 3.37M |
| 26 | DPA3-v1-OpenLAM | OMat24+ MPtrj+sAlex+Alex | 113M | 8.18M |

# 3 Results

## 3.1 Datasets collection and analysis

We have generated and collected extensive DFT datasets for various metals and alloys containing defects, as illustrated in Fig. 1. The details of these datasets are summarized in Table 2. First, we consider simple GBs for 56 elements (the GB-56 dataset), covering nearly all metals in the periodic table, as shown in Fig. 1(a). These GB structures, obtained from a previous study [40], include relaxed GB structures as well as random lattice perturbations of 1% and 2% of the lattice constant. These perturbations are introduced to evaluate the reliability of uMLIPs for non-equilibrium GBs. In total, this dataset comprises 1,436 structures. Second, we consider four BCC refractory metals: Mo, Nb, Ta, and W. We adhere to the protocol from our prior study to generate comprehensive defect genomes [41]. These genomes are combined with a standard dataset constructed using domain knowledge, which includes ground-state structures, configurations deformed under various elastic strains, ab initio molecular dynamics (AIMD) simulations at multiple temperatures, edge and screw dislocations, and simple grain boundaries. This integration results in the creation of comprehensive datasets: Mo-g, Nb-g, Ta-g, and W-g. These datasets are designed for general-purpose applications, with a representative configuration shown in Fig. 1(b). It should be noted that these datasets incorporate all the possible atomic environments that emerge during extensive plastic deformations in BCC metals, encompassing processes such as polycrystalline compression and tension, single crystal compression and tension, nanoindentation, and crack propagation [41]. To evaluate the performance of uMLIPs on systems with free surfaces, we also include the W-s dataset, generated through embedded atom model-guided sampling in our prior work [41]. This dataset consists of clusters containing defect derived from general GBs in random polycrystals, as depicted in Fig. 1(c). Additionally, for Mg, we utilize the RANDSPG dataset from Ref. [42], which includes a diverse range of structures encompassing all crystal space groups, as illustrated in Fig. 1(d). This dataset is referred to as RANDSPG throughout this work.

For systems containing multiple elements, we consider the following datasets: $Mo_{50}Nb_{50}$-d, $Mo_{25}Nb_{25}Ta_{25}W_{25}$-d, $Mo_{25}Nb_{25}Ta_{25}W_{25}$-H, CrCoNi, HEA10, and MoNbTaW, which spans the entire compositional space. For the $Mo_{50}Nb_{50}$-d and $Mo_{25}Nb_{25}Ta_{25}W_{25}$-d datasets, "d" represent defect genomes of equimolar alloys, generated by rescaling lattice constants and substituting elements within the W defect dataset [41]. The $Mo_{25}Nb_{25}Ta_{25}W_{25}$-H dataset includes the diffusion of a single atomic hydrogen around a screw dislocation core (Fig. 1(e)). The CrCoNi dataset (Fig. 1(f)), sourced from a recent publication [43], was used to study the formation of chemical short-range order of the medium-entropy CrCoNi alloy. Additionally, we generate a small DFT dataset HEA10 (Fig. 1(g)), containing 10 elements Al, Hf, Mo, Nb, Ni, Ta, Ti, V, W, Zr, which is used to study the performance of uMLIPs on such complex systems with many elements. Additionally, we consider a comprehensive dataset developed for MoNbTaW spanning the entire compositional space. This dataset was used to train sMLIPs in the atomistic simulation of dislocation motion and polycrystal compression in our recent work [44]. Finally, to directly validate the accuracy of solute-defect interactions in alloys, we utilize data from previous study [45]. This dataset comprises solute-defect interaction energies in Mo and Ta matrices, with Pt, Re, Ta, Os, and Hf as solute elements. It includes a variety of defects, such as monovacancies, dumbbell vacancies, different types of GBs, generalized stacking faults, and screw dislocations.

The energies and atomic forces of all datasets derived from DFT calculations are presented in Fig. 1(h-i). These datasets encompass a total of 54,084 configurations, a significant portion of which exhibit positive energies. These positive-energy configurations primarily correspond to the RANDSPG dataset for Mg. The broad distribution of energies, ranging from highly stable to highly unstable configurations, demonstrates that our study captures a comprehensive spectrum of chemical and structural complexity. Furthermore, Fig. 1(i) illustrates the distribution of atomic forces across a total of 7,326,684 atoms. The force magnitudes exhibit a wide range, reflecting the varied atomic environments present in the dataset, from bulk-like regions to highly distorted configurations near defects and interfaces. This thorough coverage ensures that the conclusion of this study is not only applicable for low-energy, stable configurations but also capable of handling high-energy, metastable, and defect-rich environments.

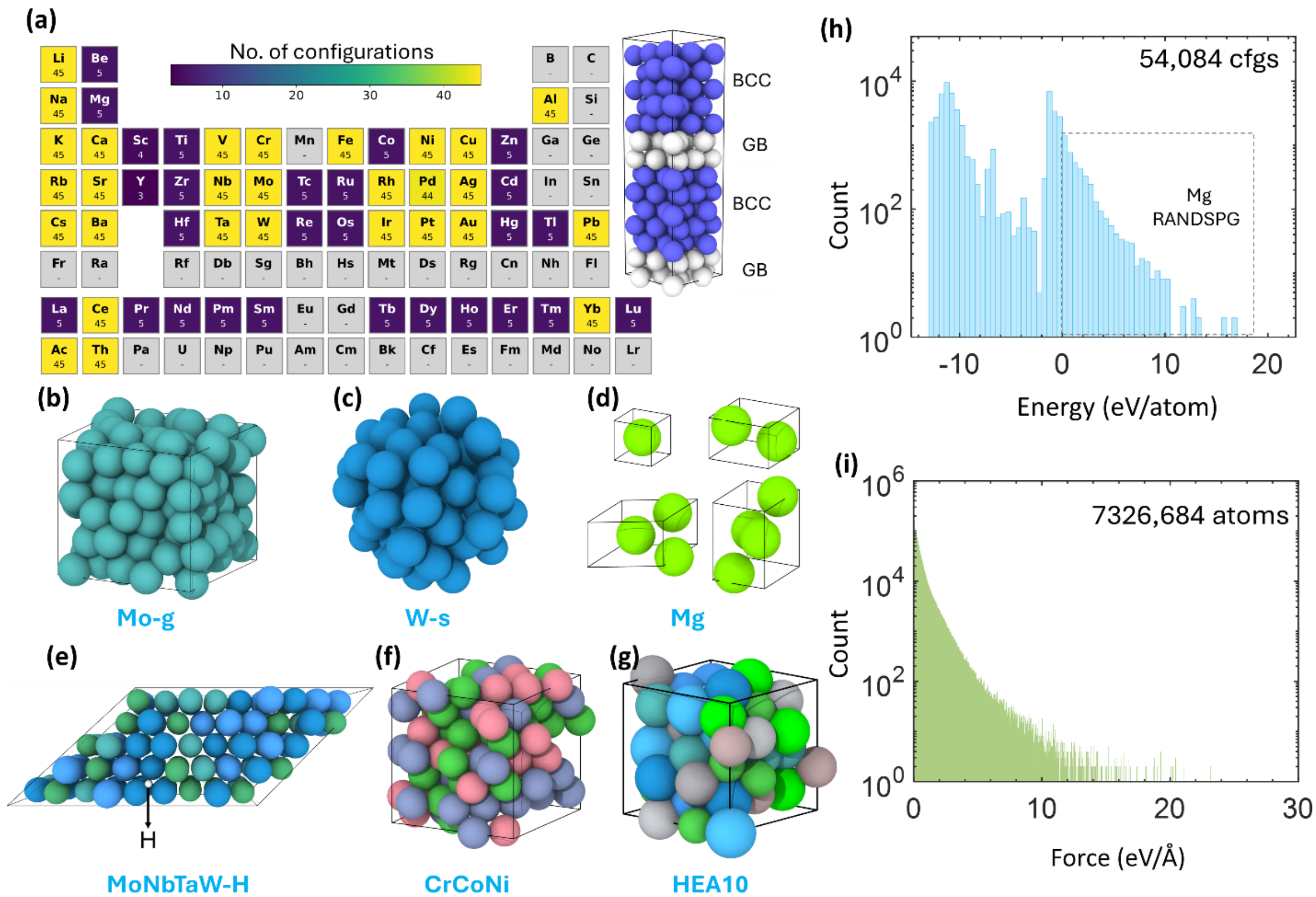

Figure 1 DFT datasets used for the assessment of uMLIPs. (a) Simple GBs for 56 elements, with the corresponding structure counts for each element. A BCC bicrystal containing two GB are presented. (b) BCC Mo-g. "g" here denotes "general-purpose". (c) BCC W-s. "s" here denotes "spherical". (d) Mg dataset obtained by RANDSPG. (e) MoNbTaW-H dataset. A single atomic H is near a screw dislocation core. (f) CrCoNi dataset. (g) HEA10 containing 10 elements i.e. Al, Hf, Mo, Nb, Ni, Ta, Ti, V, W, Zr. (h) Histogram plot of energy for all datasets. The dashed rectangular box highlights the source of positive energy. (i) Histogram plot of atomic force for all datasets.

It is important to note that the datasets listed in Table 2, which are used to benchmark the uMLIPs in this study, differ significantly from those employed in previous research [5,29–31,46] since the existing studies have predominantly focused on perfect crystalline structures or surfaces, which do not fully capture the realistic but complex microstructural features and interactions. In contrast, defects—such as vacancies, dislocations, and grain boundaries—are crucial determinants of the mechanical, thermal, and chemical properties of metals and alloys [47]. Our datasets, illustrated in Fig. 1 and detailed in Table 1, encompass the most comprehensive defects for pure metals, as well as intricate interactions among various chemical elements in multi-component systems. By including a wide range of defect types and chemical environments, these datasets provide a robust foundation for validating the accuracy and generalizability of uMLIPs in modeling material behavior under diverse and challenging practical conditions.

Table 2 DFT datasets used to assess the accuracy of uMLIPs.

| ID | Dataset | Information | cfgs |
|---|---|---|---|
| 1 | GB-56 | All relaxed and rattled GBs of 56 metals (this work) | 1436 |
| 2 | Mo-g | Domain knowledge & extensive defects (this work) | 1030 |
| 3 | Nb-g | Domain knowledge & extensive defects (this work) | 1018 |
| 4 | Ta-g | Domain knowledge & extensive defects (this work) | 1050 |
| 5 | W-g | Domain knowledge & extensive defects [41] | 1026 |
| 6 | W-s | Spherical clusters containing defect [41] | 262 |
| 7 | Mg | RANDSPG [42] | 17210 |
| 8 | $Mo_{50}Nb_{50}$-d | Extensive defects (this work) | 347 |
| 9 | $Mo_{25}Nb_{25}Ta_{25}W_{25}$-d | Extensive defects (this work) | 348 |
| 10 | $Mo_{25}Nb_{25}Ta_{25}W_{25}$-H | Atomic H around screw dislocation core (this work) | 998 |
| 11 | CrCoNi | Domain knowledge with magnetic calculations [43] | 1257 |
| 12 | HEA10 | Bulk AIMD: Al, Hf, Mo, Nb, Ni, Ta, Ti, V, W, Zr (our work) | 600 |
| 13 | MoNbTaW | All compositional space [44] | 17654 |
| 14 | Solute-defects interactions | Mo-Pt, Mo-Re, Mo-Ta, Ta-Os, Ta-Hf [45] | 409 |

## 3.2 Accuracy assessment across different uMLIPs and datasets

We begin by computing the energies and forces of the GB-56 dataset using both DFT and all the uMLIPs under consideration in this study. The root mean square errors (RMSE) for energy and force are then calculated for each element. Fig. 2 presents the results for CHGNet and eqV2-31M-omat-mp-salex, representing the least and most accurate models, respectively. CHGNet demonstrates good performance for metals in groups 1-3, with energy RMSE below 10 meV/atom and force RMSE below 50 meV/Å. However, for transition metals in groups 4-11, CHGNet exhibits significant errors, with energy RMSE ranging from 10 to 37 meV/atom and force RMSE ranging from 100 to 455 meV/Å. For metals in groups 12-14, noticeable errors are observed for Al and Pb. Among the lanthanoids and actinoids, Ce, Yb, Ac, and Th show large energy and force RMSE values. In contrast, the eqV2-31M-omat-mp-salex model significantly outperforms CHGNet, as evidenced by the narrower range of the color bars for energy and force RMSE in Fig. 2. The eqV2-31M-omat-mp-salex model registers the highest errors, with an energy RMSE of 6.7 meV/atom and a force RMSE of 193 meV/Å. Among the 56 elements, nearly all exhibit energy RMSEs below 5 meV/atom and force RMSEs under 100 meV/Å, with only a few exceptions. Specifically, K, Fe, Ni, and W show slightly higher energy RMSE values, while Cr and Ni have force RMSE values exceeding 100 meV/Å, likely due to their complex magnetic behavior. These results highlight that while previous uMLIPs like CHGNet exhibit notable errors, particularly for transition metals and certain lanthanoids/actinoids, eqV2-31M-omat-mp-salex demonstrate remarkable accuracy for modeling simple GBs across all metal elements, approaching the precision of DFT calculations.

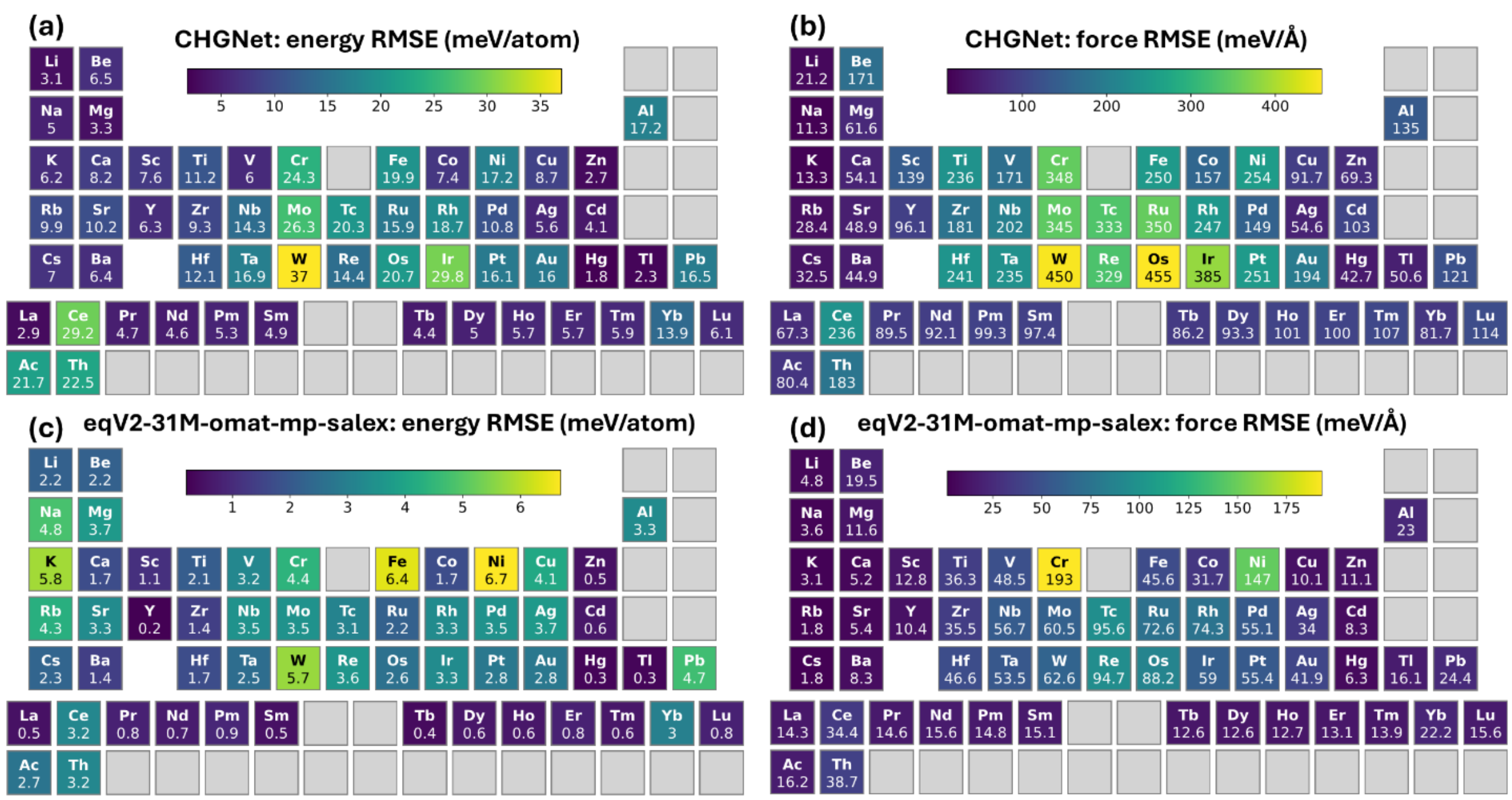

Figure 2. Performance comparison of uMLIPs on the GB-56 grain boundary dataset. (a, b) Energy and force RMSE (root-mean-square error) predictions using CHGNet; (c, d) corresponding predictions using eqV2-31M-omat-mp-salex. Numbers in the periodic table show element-specific average RMSE, with color gradients (dark purple to yellow) indicating low to high error magnitudes.

Next, we summarize the energy and force RMSE for all the datasets as presented in Table 2 and illustrate the best-performing uMLIPs for each dataset in Fig. 3. Detailed results across all datasets and uMLIPs are provided in the Supplementary Material. It is seen that the datasets GB-56, Mo-g, Nb-g, Ta-g, W-g, MoNb-d, MoNbTaW-g, MoNbTaWH, HEA10, and MoNbTaW demonstrate remarkably accurate predictions by uMLIPs, with energy RMSEs at or below 5 meV/atom and force RMSEs at or below 70 meV/Å. Notably, most datasets are best predicted by the eqV2-omat-mp-salex models for energy, with exceptions such as Nb-g, MoNbTaWH and HEA10, which are optimally predicted by the GRACE-2L-OAM, and eqV2-omat models. For force predictions, most datasets are best predicted by the eqV2-omat models, except for Mo-g, MoNb-d, MoNbTaW-H, and MoNbTaW, which are optimally predicted by the eqV2-omat-mp-salex models. The exceptional performance of eqV2 models across a wide range of datasets—spanning pure metals, binary alloys, high-entropy alloys, and complex defect structures—underscores their potential as a universal tool for high-accuracy materials modeling.

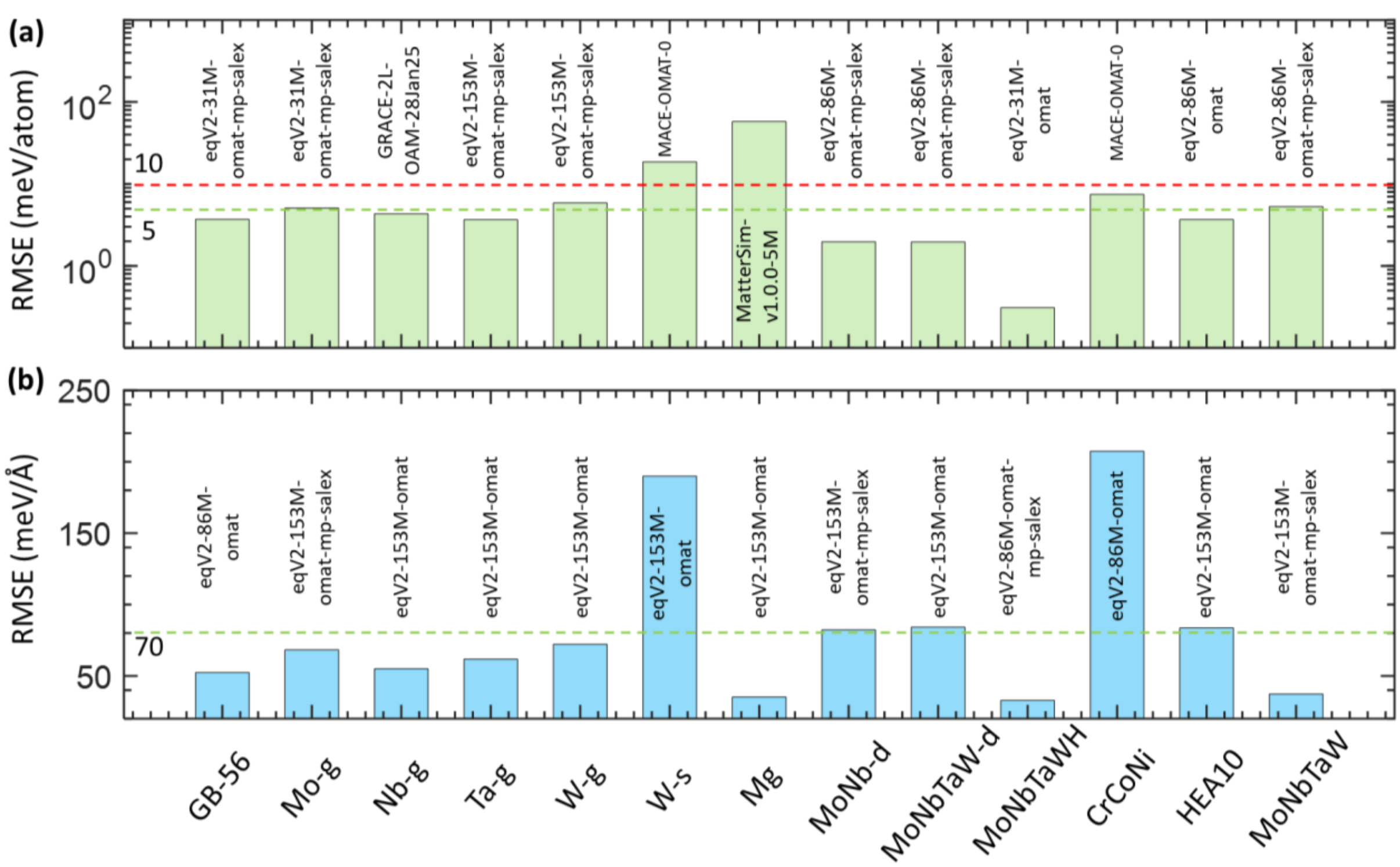

Figure 3. Comparison of the best-performing uMLIP across all benchmark datasets, showing (a) energy RMSE and (b) force RMSE. In each panel, the best-performing uMLIP is indicated on the bar for each dataset. The two dashed lines in (a) indicate energy RMSE values of 5 and 10 meV/atom, while the dashed line in (b) indicates a force RMSE value of 70 meV/atom.

Moreover, for the W-s dataset, which includes atomic clusters with free surfaces, all uMLIPs exhibit suboptimal performance. The best energy RMSE of 18.57 meV/atom is achieved by the MACE-omat-0 model, while the best force RMSE of 189.93 meV/Å is obtained by the eqV2-153M-omat model. Notably, CHGNet fails to calculate this dataset due to errors in handling isolated atoms, while the MatterSim-v1.0.0-1M model shows the highest energy RMSE of 869 meV/atom and the highest force RMSE of 1.37 eV/Å. These results highlight that atomic clusters with free surfaces remain a significant challenge for uMLIPs. While all models struggle to achieve high accuracy for this dataset, their performance varies considerably. MatterSim performs the worst, while MACE and eqV2 models demonstrate the best performance among the tested uMLIPs. This variability underscores the need for further development of uMLIP architectures and training strategies to better handle systems with free surfaces.

For the Mg dataset, MatterSim-v1.0.0-5M demonstrates the best energy predictions, achieving an energy RMSE of 57.80 meV/atom and force RMSE of 41.11 meV/Å, while eqV2-153M-omat leads in force prediction with a force RMSE of 35.24 meV/Å. Notably, MatterSim is the only model that achieves good accuracy for both energy and force predictions on this dataset, which features configurations indicative of very high energy. This capability stems from MatterSim's training on 17 million configurations spanning high-temperature and high-pressure ranges, making it uniquely suited for such extreme conditions. Fig. 4 presents a parity plot of predicted energies by uMLIPs compared to the reference energies obtained from DFT calculations. The plot reveals that CHGNet and eqV2-31M-omat-mp-salex significantly underestimate positive energies, underscoring MatterSim's superior performance in high-pressure and high-temperature conditions.

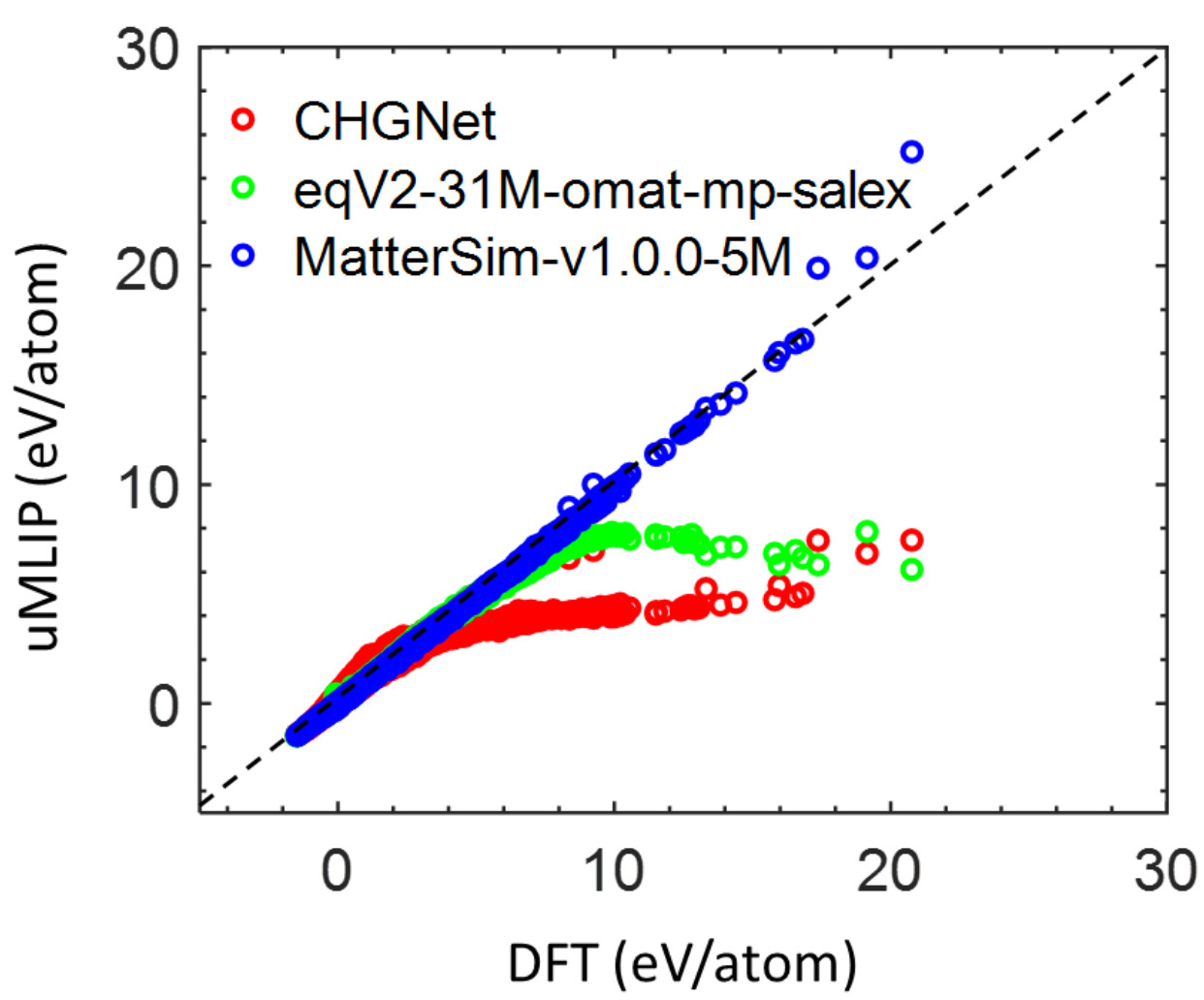

Figure 4 Parity plot of uMLIP-predicted energies versus DFT reference energies for Mg structures in the RANDSPG dataset. The dataset contains many high-energy configurations with large positive values (Fig. 1(h)), which explains the poorer performance of the eqV2 and CHGNet potentials in this regime.

For the CrCoNi dataset, which includes rattled FCC lattices and liquid configurations of equimolar medium entropy CrCoNi alloys, the MACE-OMAT-0 model delivers highly accurate energy predictions, achieving an energy RMSE of 7.45 meV per atom. This outperforms the eqV2 86M omat model, which has an energy RMSE of 8.04 meV per atom. On the other hand, the eqV2 86M omat model provides the best force predictions, with an RMSE of 207.35 meV per Å. Given the complex magnetic distributions in random CrCoNi alloys and the inclusion of liquid configurations, these low error values—particularly for energy predictions—are highly noteworthy. It is also important to note that the CrCoNi dataset includes crystalline phases with chemical ordering, extracted from various points along DFT Monte Carlo simulations. The exceptional performance of the MACE-OMAT-0 model suggests that it is particularly well suited for studying CrCoNi alloys, such as investigating the formation mechanisms of chemical short range order (CSRO), without the need for expensive DFT calculations.

We next rank all uMLIPs based on their energy and force prediction capabilities. To ensure a fair comparison, we exclude the W-s dataset, as all uMLIPs struggle with clusters with free surfaces. The RANDSPG dataset for Mg is omitted because only MatterSim achieves moderate accuracy for its energy predictions. The CrCoNi dataset is not considered due to its inclusion of liquid configurations. RMSE values for energy and force were computed over the remaining combined datasets. The uMLIPs trained exclusively on the MPtrj dataset are represented in red. Fig. 5 presents the energy RMSE and force RMSE plotted against the different uMLIPs. In terms of energy predictions (Fig. 5(a)), the eqV2-omat-mp-salex models rank as the top three uMLIPs, demonstrating superior accuracy. Securing fourth and fifth places are GRACE-2L-OAM and DPA3-v1-OpenLAM, respectively, followed by GRACE-1L-OAM, MACE-MPA-0, MatterSim-v1.0.0-5M, eqV2-86M-omat, and eqV2-31M-omat. Interestingly, eqV2-31M-mp ranks next, outperforming MACE-omat-0, MatterSim-v1.0.0-5M, DPA3-v1-MPtrj, and eqV2-153M-omat. The subsequent ranking order includes Orb-MPtraj-only-v2, Orb-d3-v2, Orb-d3-sm-v2, GRACE-2L-r6-MP, Orb-d3-xs-v2, GRACE-1L-r6-MP, MACE-MP-0, and CHGNet. Notably, eqV2 models employing DNeS and trained solely with the MPtrj dataset perform the worst, suggesting that the DNeS technique may negatively impact model accuracy for energy prediction.

For force predictions (Fig. 5(b)), the six eqV2 models trained with the OMat24 dataset are the most accurate uMLIPs, with force RMSE values only half those of other models. This exceptional accuracy stems from OMat24's approximately 110 million non-equilibrium configurations, which are essential for predicting atomic forces in diverse systems. Following the eqV2 models, DPA3-v1-OpenLAM, GRACE-2L-OAM, MACE-omat-0 rank, and GRACE-1L-OAM as the next most accurate models, though it is significantly less accurate than the eqV2-omat models. This gap highlights the importance of both architecture and parameter count in determining model accuracy. The remaining uMLIPs are ranked as follows: MACE-MPA-0 > Orb-d3-v2 > MatterSim-v1.0.0-5M > eqV2-dens-86M-mp > Orb-d3-sm-v2 > eqV2-dens-153M-mp > eqV2-dens-31M-mp > Orb-d3-xs-v2 > DPA3-v1-MPtrj > eqV2-31M-mp > GRACE-2L-r6-MP > MatterSim-v1.0.0-1M > Orb-MPtraj-only-v2 > GRACE-1L-r6-MP > CHGNet > MACE-MP-0. This ranking underscores the critical role of dataset quality, model architecture, and parameter count in achieving high accuracy for force predictions, with eqV2 models emerging as the state of the art.

To elucidate the impact of training datasets and architecture, we compare uMLIPs trained on the same datasets. For example, uMLIPs trained exclusively with the MPtraj dataset show the following energy prediction hierarchy: eqV2-31M-mp > DPA3-v1-MPtrj > Orb-MPtraj-only-v2 > GRACE-2L-r6-MP > GRACE-1L-r6-MP > MACE-MP-0 > CHGNet. Additionally, MACE-MPA-0 (trained on MPtraj and Alexandria-related datasets) significantly outperforms Orb models in energy predictions. Similarly, MACE-omat-0 is comparable to eqV2-omat models despite having far fewer parameters (9.06M vs. 31/86/154M). For force predictions, the inclusion of OMat24 consistently improves accuracy, while uMLIPs trained solely on MPtraj perform the worst. Although the GRACE-2L-OAM and DPA3-v1-OpenLAM model incorporate 113 million structures in its training dataset, its performance in predicting energy and force is less accurate than the eqV2 models. This disparity is likely due to the insufficient number of parameters in these two models. Overall, while it is challenging to definitively conclude which uMLIP is the best due to varying datasets and parameter counts, GRACE-2L, DPA3-v1 and eqV2 models trained by "OMat24+MPtrj+sAlex" stand out as top performers. These results highlight the importance of diverse datasets for energy predictions and the critical role of OMat24 for force predictions.

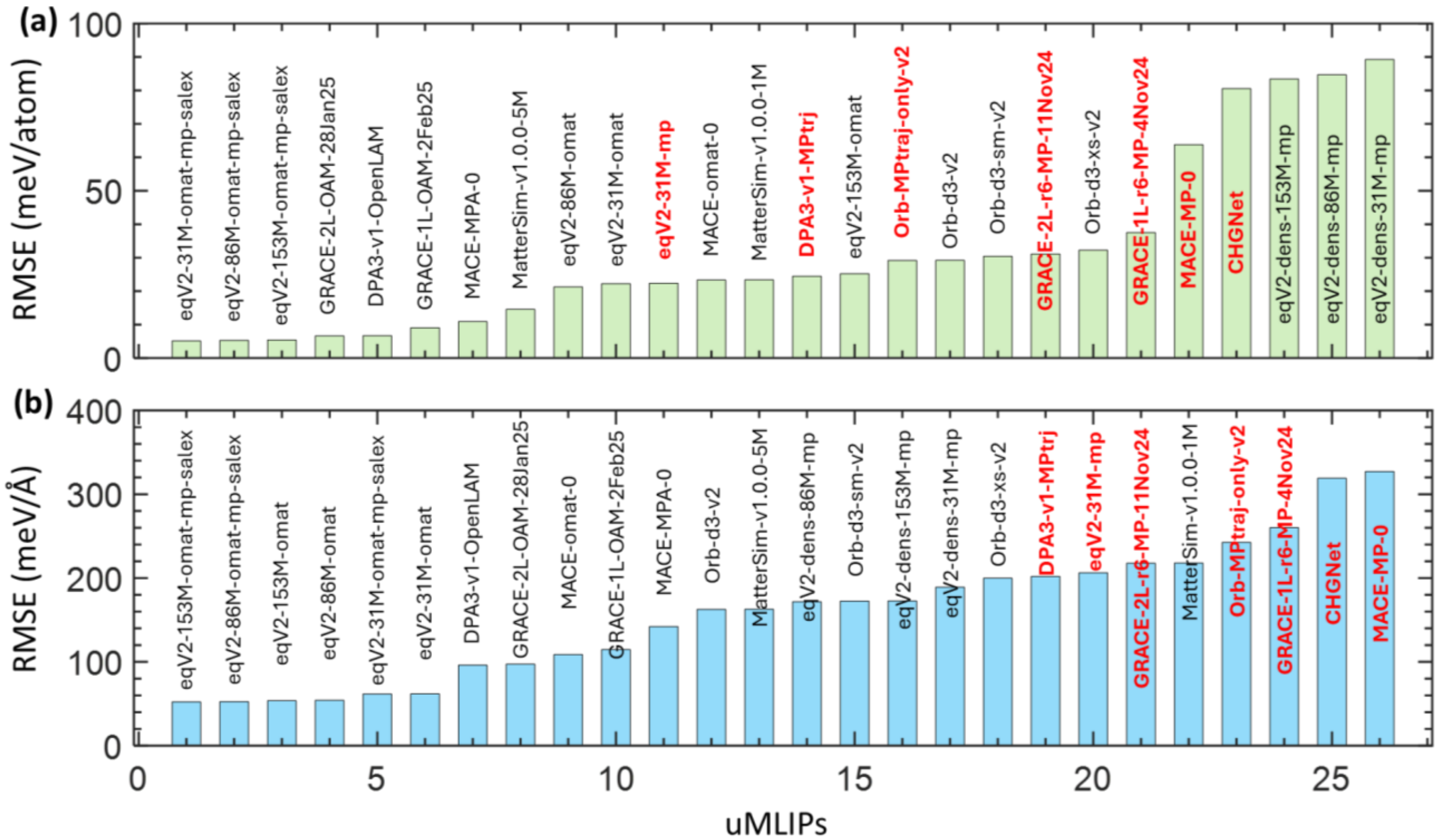

Figure 5. Ranking of uMLIP models based on (a) energy RMSE and (b) force RMSE. Models trained exclusively on the MPtrj dataset are shown in red. RMSE values for energy and force were computed over the combined datasets (excluding W-s, RANDSPG-Mg, and CrCoNi). Model names appear above each bar in ranking order.

Finally, we validate the performance of the eqV2-31M-omat-mp-salex model, one of the most accurate uMLIPs, in simulating solute-defect interactions in BCC metals. The DFT reference data for solute-defect interaction energies are taken from a previous study [45], while the corresponding predictions are computed using the eqV2-31M-omat-mp-salex model with the same configurations. The results, presented in Fig. 6, cover five substitutional-defect systems: W-Pt, W-Re, W-Ta, Ta-Os, and Ta-Hf. We consider 11 common defects, each evaluated at multiple data points by varying the distance between the solute atom and the defects. It is observed that the predictions by eqV2-31M-omat-mp-salex closely approximate the DFT values across all defect types in all systems. Prediction errors are quantified by RMSE values, ranging from 40.2 meV to 84.4 meV. These low error values demonstrate that the eqV2-31M-omat-mp-salex model achieves near-DFT accuracy in modeling solute-defect interactions in these systems.

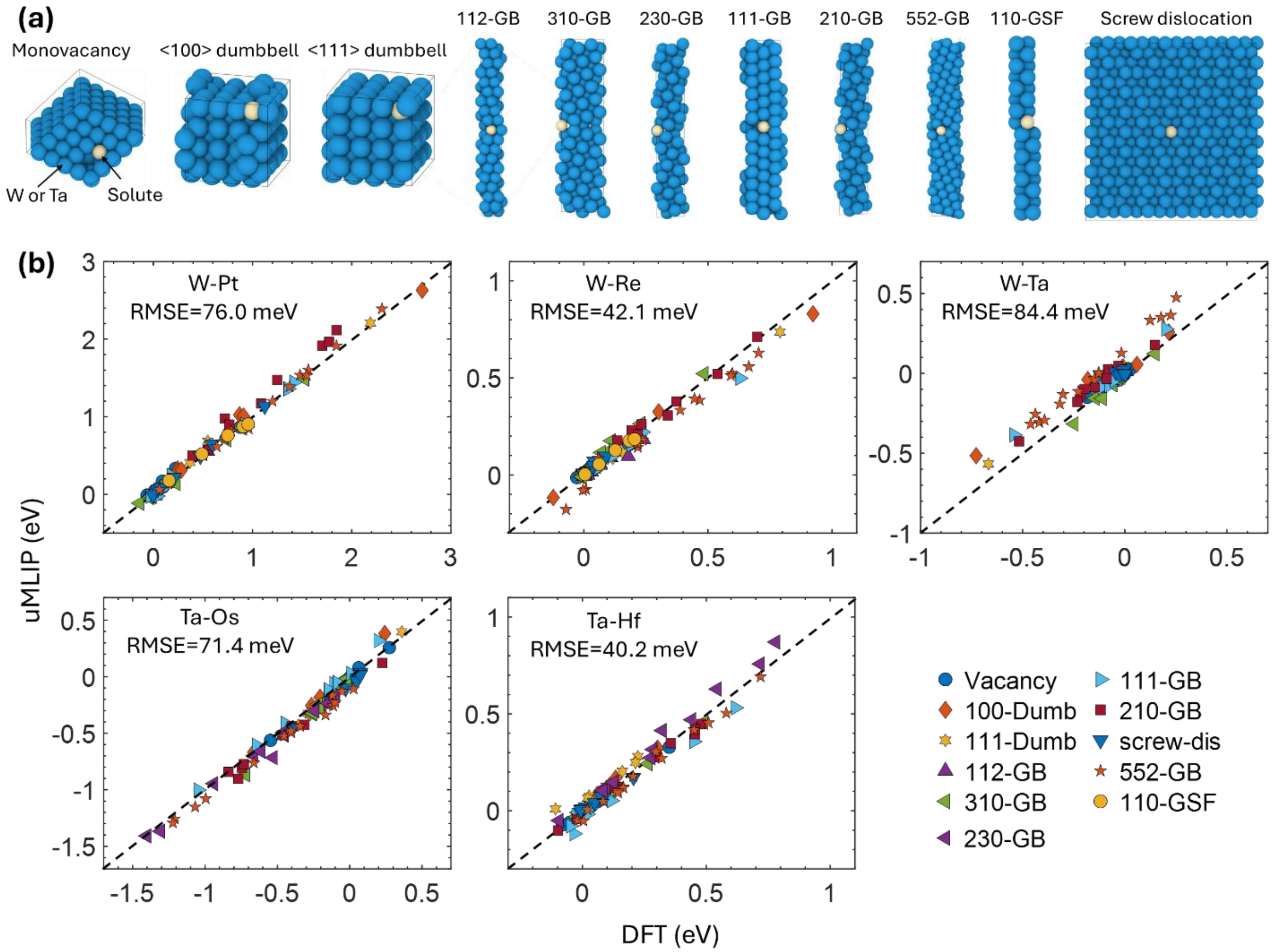

Figure 6. Validation of eqV2-31M-omat-mp-salex for solute-defect interaction energy in bcc metals. (a) Atomic configurations of defect types studied: monovacancy, <110> and <111> dumbbell interstitials, seven GBs, a generalized stacking fault (GSF), and a screw dislocation. (b) Parity plot comparing DFT-calculated versus uMLIP-predicted solute-defect interaction energies for W-Pt, W-Re, W-Ta, Ta-Os, and Ta-Hf systems. The dashed line denotes perfect agreement, and the RMSE for each system is indicated.

### 3.3 Trade-off between computational accuracy and cost

We conduct a systematic analysis to assess the computational accuracy versus cost for several models: EAM, sMLIPs, all uMLIPs, and DFT. The measurements are taken on a single core of an AMD 9654 CPU and one Nvidia A100 GPU of the Snellius supercomputer, respectively. Notably, eqV2 models with 86 M and 153 M parameters are excluded from this analysis due to their excessive computational memory demands. To evaluate the computational cost, we perform ten steps of (micro-canonical) NVE MD simulations on a 125-atom BCC W and calculate the cost per

MD step per atom. The computational accuracy is quantified by the energy and force RMSE from Fig. 5. For EAM-Zhou [48], ACE and MTPs [41], we consider only the W-g dataset, as these potentials are not universal. Both MTPs and ACE were trained exclusively on the W-g dataset from our recent study [41]. Fig. 7(a) reveals that EAM is the fastest potential with an accuracy of approximately 52.5 meV/atom, while sMLIPs such as ACE and MTPs are 1-2 orders of magnitude slower but offer accuracy approaching that of DFT. uMLIPs without fine-tuning display a significant variation in accuracy and efficiency, being 1-3 orders of magnitude slower than sMLIPs. The most accurate model, eqV2-31M-omat-mp-salex, approaches the accuracy of sMLIPs but is still 3-4 orders of magnitude faster than DFT calculations. Notably, Fig. 7(b) shows that GRACE-1L-OAM, GRACE-2L-OAM and eqV2-31M-omat-mp-salex are positioned on the Pareto frontier. eqV2-31M-omat-mp-salex, while being the least efficient, offers the highest accuracy, with its computational cost being 81 times that of the fastest GRACE-1L-OAM. Among all uMLIPs, CHGNet and MACE-MP-0 are positioned far from the Pareto frontier, indicating a less favorable balance between computational cost and accuracy.

All uMLIPs employ graph-based architectures that exploit GPU acceleration. Fig. 7(c) and 7(d) report energy and force RMSE as functions of inference cost on a single NVIDIA A100 GPU, demonstrating up to a 100× reduction in compute time compared to CPU benchmarks in Fig. 7(b). On the GPU, Orb models achieve the fastest inference speeds, followed by GRACE, MACE, DPA3, MatterSim, and the eqV2 series. Notably, Orb-d3-xs-v2, GRACE-2L-OAM, DPA3-v1-OpenLAM, and eqV2-31M-omat-mp-salex all lie on the Pareto frontier, illustrating optimal trade-offs between accuracy and cost. Moreover, the trends in Fig. 7(d) confirm that more computationally intensive uMLIPs deliver increasingly accurate force predictions. Inference speeds across the uMLIPs decrease in the sequence: Orb > GRACE > MACE > eqV2. The observed performance inversion in Fig.7(c, d), where the MatterSim-v1.0.0-5M model outperforms its 1M counterpart, presents an unexpected scaling behavior. While differences in GPU utilization patterns could potentially explain this result, particularly through the 5M model's capacity to amortize kernel launch overhead and saturate tensor core capacity via larger tensor operations, alternative explanations remain plausible and warrant systematic investigation.

Our results presented in Fig. 7 highlight the significant computational cost advantage of uMLIPs over DFT, with uMLIPs being at least three orders of magnitude faster. This advantage is even more pronounced in magnetic systems, where DFT typically requires more electronic steps.

Additionally, uMLIPs exhibit linear scaling with an increasing number of atoms [46], in contrast to the cubic scaling behavior of DFT. This indicates that the computational efficiency advantage of uMLIPs over DFT becomes increasingly significant as the number of atoms increase. Consequently, uMLIPs are well-suited for relatively large systems, capable of handling hundreds of thousands of atoms effectively.

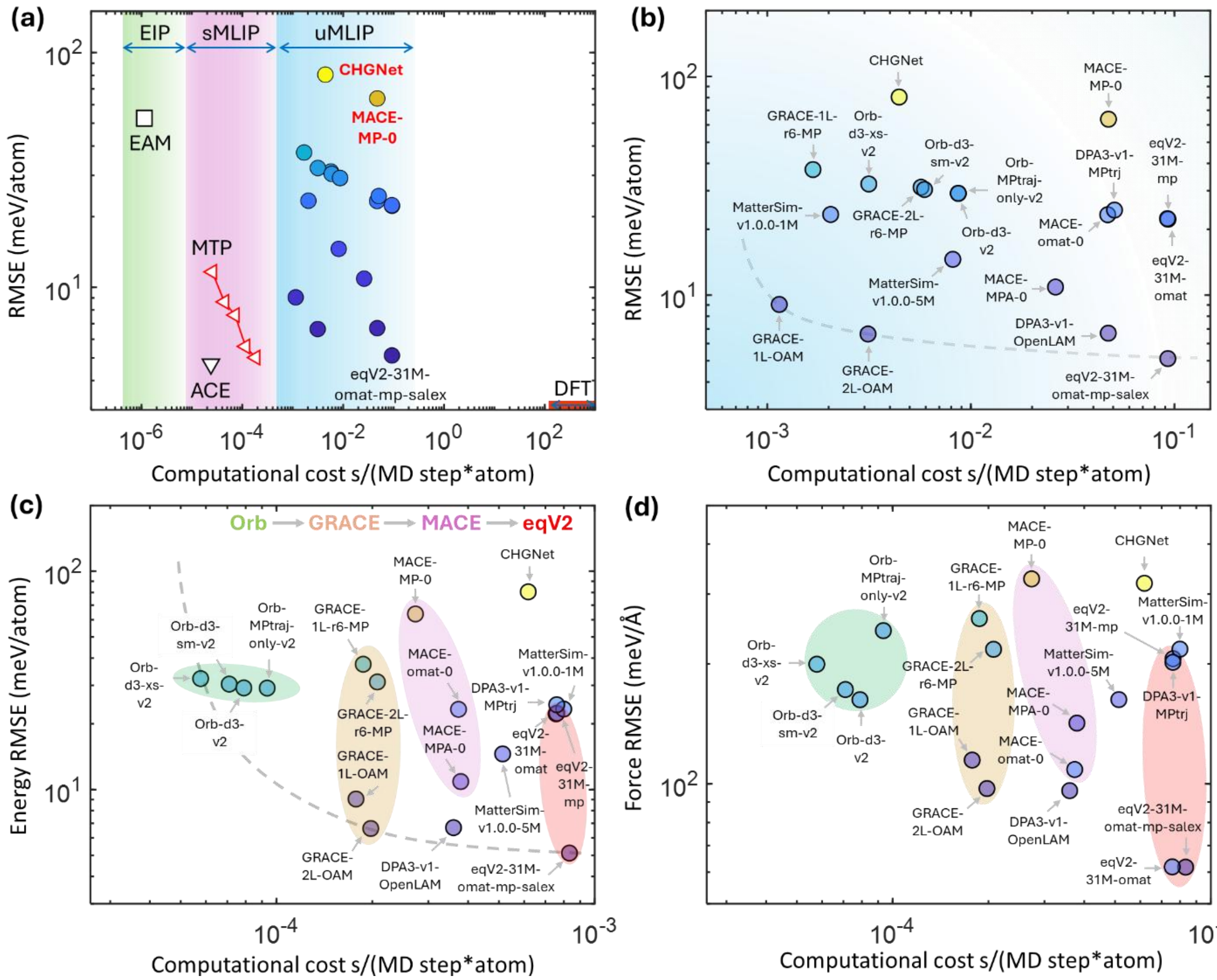

Figure 7. Accuracy–efficiency trade-off of interatomic potentials. (a) Performance on a single CPU core: classical EAM, sMLIPs (ACE and MTPs, levels 12–20), uMLIPs, and DFT. The sMLIPs (from Ref. [41]) were trained exclusively on the W-g dataset. Efficiency is reported as time per MD step per atom; accuracy is energy RMSE. (b) Detailed CPU benchmarking of uMLIPs. (c, d) GPU performance of uMLIPs in terms of energy and force RMSE. The measurements are taken on a single core of an AMD 9654 CPU and one Nvidia A100 GPU, respectively.

### 3.4 Uncertainty quantification of eqV2 models

Uncertainty quantification (UQ) is critical for the reliable application of uMLIPs. For uMLIPs to serve as a viable replacement for DFT, it is essential to ensure that predictions made by these models exhibit low uncertainty or error. However, robust UQ functionality is not inherently available for uMLIPs. To address this challenge, we adopt an ensemble strategy using six eqV2 models (models 15–20 in Table 1) to quantify predictive uncertainty. Following the Ref. [49], we compute the maximum deviation of configurational energies and atomic forces, which serve as quantitative measures of uncertainty for the whole configuration and each atom, respectively:

$$\mathrm{dev}(E) = \max_k \left| E^k - \langle E \rangle \right| \tag{1}$$

$$\mathrm{dev}(F_i) = \max_k \left| \boldsymbol{F}_i^k - \langle \boldsymbol{F}_i \rangle \right| \tag{2}$$

where $k = 1, \ldots, 6$ indexes the eqV2 models in the ensemble, $E^k$ is the energy predicted by model k, and $\langle E \rangle$ is the ensemble average of the energy. The force on atom $i$ in ensemble $k$ is given by $\boldsymbol{F}_i^k$, while $\langle \boldsymbol{F}_i \rangle$ is the ensemble force average.

We compare the maximum deviations of the energy and force, $\mathrm{dev}(E)$ and $\mathrm{dev}(F_i)$, to their respective ground-truth errors, $|E^{DFT} - \langle E \rangle|$ and $\left|\boldsymbol{F}_i^{DFT} - \langle \boldsymbol{F}_i \rangle\right|$, as shown in Fig. 8. Fig. 8(a) demonstrates that the maximum deviation in energy predictions closely tracks the RMSE across all datasets, yielding a Spearman's rank correlation coefficient of 0.88. Notably, the Mg dataset departs from the other datasets, exhibiting both larger deviations and errors, which increase the Spearman's rank correlation. Likewise, Fig. 8(b) shows that the maximum deviation in atomic forces correlates with force RMSE (Spearman's rank correlation coefficient of 0.70). By employing the maximum deviation rather than the standard deviation, we avoid underestimating uncertainty observed in MatterSim [24] and instead accentuate the spread in uMLIP predictions. Together, these findings confirm that ensemble-based estimates furnish robust and informative uncertainty quantification for all eqV2 calculations.

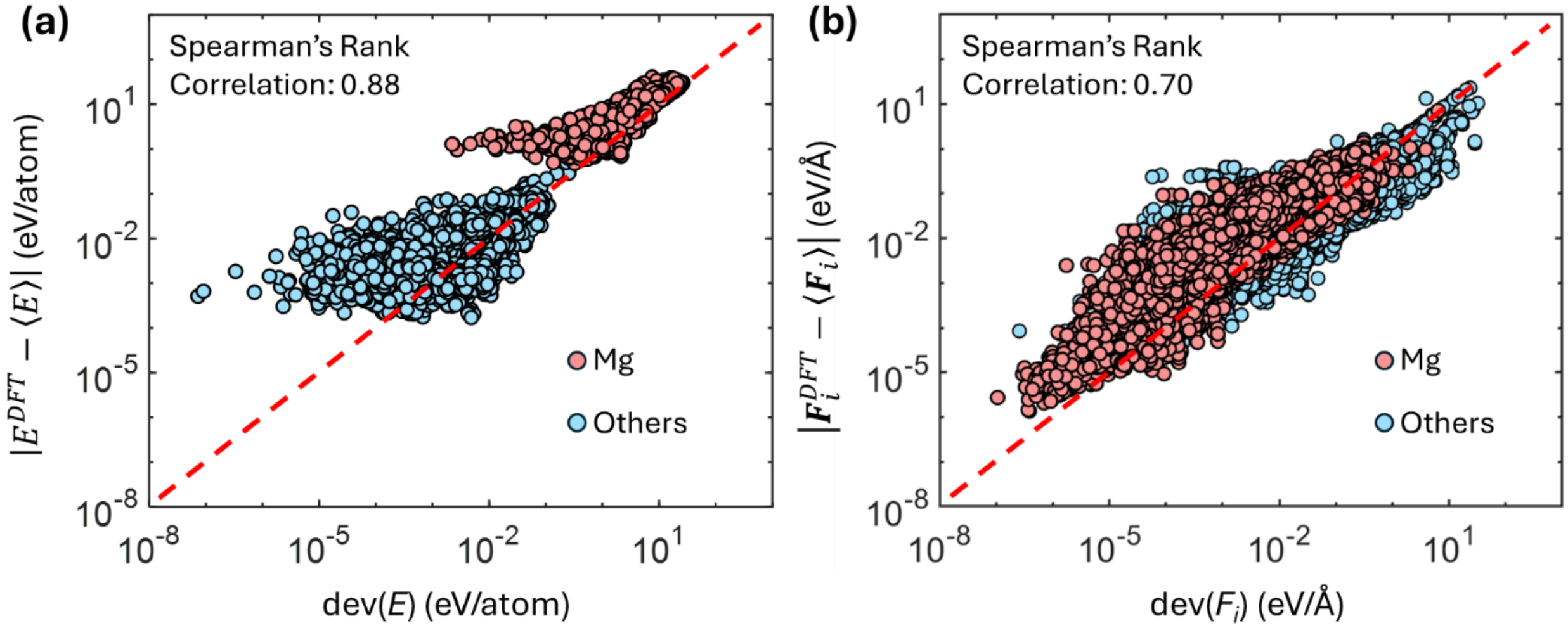

Figure 8. Ensemble-based uncertainty quantification with six eqV2 models. (a) Configurational-energy uncertainties; (b) Atomic-force uncertainties. Results for the Mg dataset are shown separately from all other datasets. The diagonal line in each panel represents perfect correlation, and the Spearman's rank correlation coefficient is reported.

### 3.5 Comparison between uMLIP and sMLIP

In this section, we compare the performance of uMLIPs and sMLIPs. Specifically, we evaluate the eqV2-31M-omat-mp-salex model against ACE and MTP in terms of accuracy. The ACE potential is specifically developed for BCC W and trained on the W-g dataset , while the MTP22 potential is trained on the MoNbTaW dataset [41]. Here, "22" refers to the level of the MTP potential, which represents a very high level of complexity suitable for practical simulations. We assess the energy RMSE and force RMSE for GBs in BCC W across various temperatures and deformation states, as well as for the entire MoNbTaW dataset. The results, displayed in Fig. 9, show that eqV2-31M-omat-mp-salex provides highly accurate predictions, with energy RMSEs between 3 and 6 meV/atom and force RMSEs between 50 and 100 meV/atom. This accuracy is maintained even under extreme conditions, such as temperatures up to 2000 K, severe plastic deformation in GBs, and complex chemical environments in MoNbTaW. These errors are significantly lower than those predicted by ACE or MTP22, particularly for force predictions, highlighting the superior performance of uMLIPs over sMLIPs in modeling complex defects and chemical interactions.

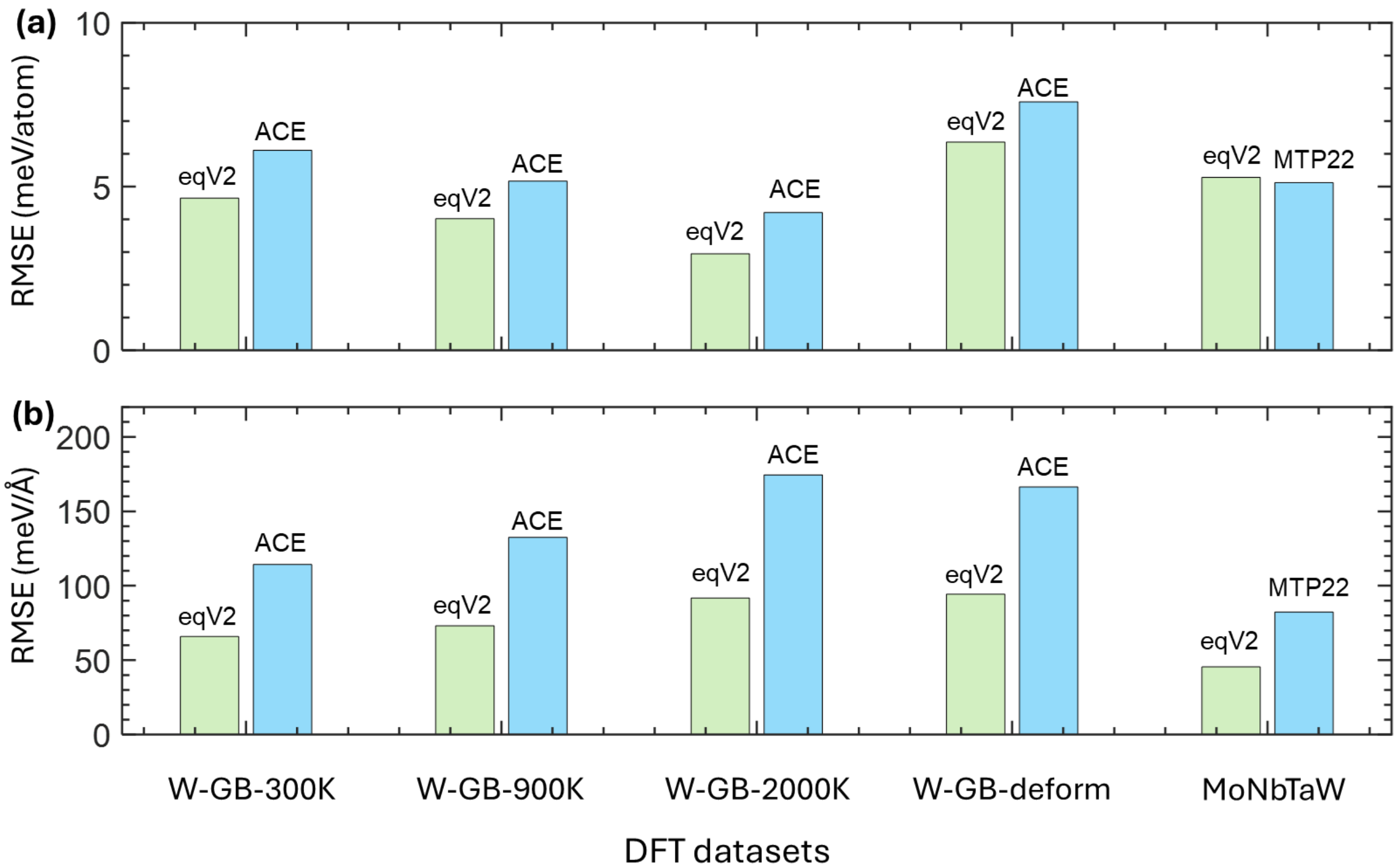

Figure 9 Comparison of prediction accuracy between eqV2-31M-omat-mp-salex and sMLIPs (ACE and MTP22) for GBs in BCC W and MoNbTaW: (a) energy RMSE and (b) force RMSE. GB prediction accuracy evaluated on four validation datasets (GB-300, GB-900, GB-2000, GB-deform) representing BCC W bicrystals at different temperatures (300 K, 900 K, 2000 K) and under severe compressive deformation [41].

It should be noted that the ACE potentials are specifically developed for W by considering extensive defects. Previous studies have shown that the extrapolation grade for general plastic deformation, based on the W-g dataset, is lower than 1. This indicates the high comprehensiveness of the W-g dataset, as it effectively captures a wide range of defect configurations and deformation scenarios. Despite this, the eqV2-31M-omat-mp-salex model outperforms ACE in accuracy, demonstrating the superior generalization capabilities of uMLIPs even when compared to highly specialized ACE. To investigate why eqV2-31M-omat-mp-salex outperforms ACE, we extract all configurations containing only W from the MPTrj, sAlex, and OMat24 datasets, totaling 40, 49, and 425 structures, respectively. These configurations are collectively referred to as the W-uMLIP dataset. We then conduct a principal component analysis (PCA) on the local atomic environments (LAEs) of each atom in these configurations, using the Smooth Overlap of Atomic Positions (SOAP) descriptor to quantify the LAEs [50]. It should be noted that PCA reduces high-dimensional data (the SOAP vector here) by extracting orthogonal principal components (PCs) that capture maximum variance. The first two components (PC1 and PC2) are most important as they typically represent the dominant data trends, enabling clear 2D visualization of complex data. The W-g dataset is also included for comparison. The first two principal components are plotted against each other in Fig. 10(a), revealing that W-g encompasses a wide range of LAEs corresponding to various defects. In contrast, MPTrj primarily includes relaxation trajectories of eight different lattice structures, while sAlex contains a broader array of structures. Interestingly, OMat24 features more complex LAEs derived from non-equilibrium configurations in MPTrj and sAlex. However, even when combined, MPTrj, sAlex, and OMat24 contain significantly fewer LAEs than W-g.

To further discern the differences between eqV2-31M-omat-mp-salex and ACE, we developed a new sMLIP, ACE-new, using the W-uMLIP dataset. Fig. 10(b) illustrates that eqV2-31M-omat-mp-salex consistently shows higher energy RMSE values than ACE-new, although the dataset W-uMLIP is included in the training processing of eqV2-31M-omat-mp-salex. Subsequently, we employ ACE-new to predict various GB datasets for W under different temperatures and deformation states, as depicted in Fig. 10(c). Interestingly, ACE-new exhibits undesirable energy RMSE values, significantly higher than those of both eqV2-31M-omat-mp-salex and ACE trained with W-g. This indicates that the W-select dataset alone is not enough for modeling complex GBs in W. This suggests that the superior accuracy of eqV2-31M-omat-mp-salex arises not only from configurations involving W but also from its ability to generalize across diverse atomic environments and interactions beyond a single element.

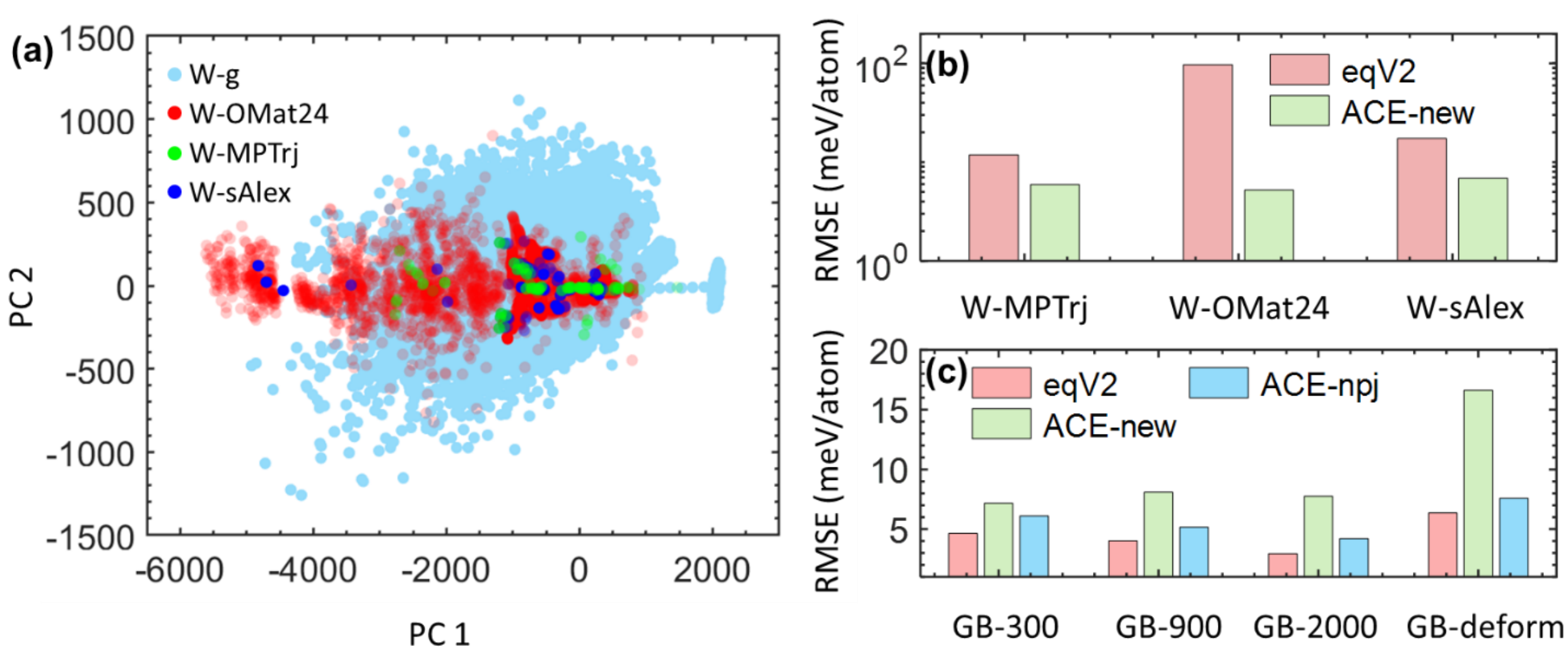

Figure 10. Comparative analysis of atomic environments and model performance. (a) Principal component analysis (PCA) of atomic environments across four W datasets: W-g and W-extracted subsets (W-OMat24, W-sAlex, W-MPTrj) from the OMat24, sAlex, and MPTrj multicomponent datasets. Points represent individual atomic configurations, colored by dataset origin. (b) Energy prediction RMSE for W structures (W-OMat24, W-sAlex, W-MPTrj) comparing the universal eqV2-31M-omat-mp-salex model with the specialized ACE-new potential trained on the combined W-extracted dataset. (c) GB energy prediction accuracy evaluated on four validation datasets (GB-300, GB-900, GB-2000, GB-deform) representing BCC W bicrystals at different temperatures (300 K, 900 K, 2000 K) and under severe compressive deformation [41]. Results compare three models: universal eqV2-31M-omat-mp-salex, specialized ACE-npj (W-g trained), and specialized ACE-new (W-extracted trained).

## 4 Discussion

We have generated and collected DFT datasets that encompass a comprehensive array of defects in metals and alloys. These datasets have been meticulously curated to cover a wide spectrum of defects, ranging from dislocations and complex grain boundaries to interstitial atoms, ensuring a robust representation of potential structural imperfections. By incorporating various alloy compositions and diverse defect configurations, our datasets serve as a valuable resource for validating the predictive capabilities of uMLIPs. Remarkably, the calculated energy and force RMSEs are below 5 meV/atom and 100 meV/Å, respectively, outperforming documented sMLIPs such as MTP and ACE in modeling complex defects. Particularly, given that the datasets Mo-g, Nb-g, Ta-g, and W-g encompass all possible atomic environments encountered in complex deformation scenarios, and considering that these elements were not specifically treated during the pretraining process, we believe that the eqV2-omat-mp-salex models can simulate defects in other metals with comparably high accuracy. In applications to random alloys, eqV2 also demonstrates excellent accuracy in systems such as $Mo_{50}Nb_{50}$-d, CrCoNi (with magnetism calculations), $Mo_{25}Nb_{25}Ta_{25}W_{25}$-d, $Mo_{25}Nb_{25}Ta_{25}W_{25}$-H, and HEA10-AlHfMoNbNiTaTiVWZr. These datasets include extensive defects, complex chemical ordering, and sophisticated elemental interactions. Additionally, we find that eqV2-31M-omat-mp-salex can capture the subtle energy changes in solute-defect interactions due to varying defect-defect distances in W and Ta binary alloys. Collectively, our results demonstrate that eqV2, trained on MPTrj, sAlex, and OMat24, can effectively replace computationally costly DFT calculations in key applications within the mechanical and materials science communities.

On the other hand, our results highlight a significant advantage of uMLIPs over sMLIPs. As demonstrated in Figs 9 and 10, uMLIPs exhibit superior extrapolation capabilities. For instance, despite the limited data for pure W in the MPTrj, sAlex, and OMat24 datasets, the eqV2-omat-mp-salex model accurately predicts complex defects in W, as demonstrated by its high accuracy on the W-g dataset. In contrast, sMLIPs such as ACE or MTP require explicit training on these defects, as their parameters are constrained to specific elements and inter-element interactions, limiting their transferability. uMLIPs, however, leverage shared parameters and element embeddings to generalize across the periodic table. Element embeddings encode the chemical identity of each atom, capturing similarities between elements based on their electronic structure, atomic size, and bonding behavior. This allows uMLIPs to learn a unified representation of atomic interactions that extends beyond the training data, enabling accurate predictions even for untrained defects or elements [21,51–54]. This flexibility reduces the need for defect-specific training data, making uMLIPs a powerful tool for modeling diverse materials and defects in these materials. One promising application is using uMLIPs to develop sMLIPs via teacher-student architectures for knowledge distillation, as uMLIPs can study large defects that DFT cannot handle. This approach has been demonstrated by the DPA-2 architecture [27], though the accuracy of DPA-2 models remains critical for success.

Finally, our results provide critical insights into the future direction of uMLIPs. A key question arises: should the focus be on developing more advanced machine learning architectures, or on generating more comprehensive DFT datasets? The ongoing competition in the Matbench-discovery repository highlights that both the academic community and industrial companies, including Meta and Microsoft, are actively pursuing both avenues. These efforts involve developing more sophisticated architectures with increased parameters and training on larger datasets. For instance, the advanced eqV2 models now contain 153 million parameters and are trained on 110 million structures, showcasing the trend toward scaling up both model complexity and dataset size. Although extensive model parameters in the eqV2 architecture remains debatable [55], our results demonstrate that eqV2 models are already highly accurate for predicting general defects in metals and alloys, with accuracy approaching the noise level of DFT calculations. This indicates that the three datasets—MPTrj, sAlex, and OMat24—are sufficient for modeling defects in bulk systems. However, the next steps should focus on addressing current limitations of uMLIPs. First, more structures involving free surfaces need to be included, as current uMLIPs still struggle with surface-dominated systems. Second, for magnetic systems such as Fe, Cr, and CrCoNi, our results indicate relatively low accuracy, suggesting the need to explicitly incorporate magnetic properties as outputs, as demonstrated by models like CHGNet. Third, datasets from MatterSim, which include high-temperature and high-pressure structures, should be integrated to improve performance in extreme-condition applications. Addressing these limitations will be crucial for expanding the applicability of uMLIPs to a broader range of materials science problems. Another key aspect for advancing uMLIPs is the development of reliable UQ methods. By quantifying uncertainty, users can identify regions of low confidence in predictions and take appropriate measures to validate or refine results. Although our results in Fig. 8 demonstrate the potential of ensemble methods in eqV2 models for UQ, further improvements are needed. This advancement is crucial for the future goal of enabling MLIPs to fully replace DFT calculations for any kind of microstructural features and defects present in wide class of materials.

Lastly, we emphasize that our findings focus specifically on static property predictions, particularly configurational energy and atomic forces. While these metrics are fundamental to materials characterization, we acknowledge the critical importance of additional properties such as energy Hessians [56], phonon spectra [5], dynamic behaviors [57], and simulation stability [58] for practical applications of uMLIPs. This limitation is particularly relevant for non-conservative models like eqV2, where forces are directly trained from raw data rather than derived as energy gradients. However, our systematic validation across all metal GBs (Fig. 2) and solute-defect systems (Fig. 6) demonstrates eqV2's robust capability in predicting key static properties such as defect-solute interaction energies, which directly govern mechanical performance metrics like strength and embrittlement resistance.

## 5 Conclusions

In conclusion, our study demonstrates the remarkable potential of uMLIPs in accurately modeling defects and complex interactions in metals and alloys, rivaling the precision of DFT at a fraction of the computational cost. The eqV2 models, trained on diverse datasets such as MPTrj, sAlex, and OMat24, achieve exceptional accuracy in predicting energies and forces, with RMSEs below 5 meV/atom and 100 meV/Å, respectively. These models outperform sMLIPs like ACE and MTP, particularly in extrapolating to unseen defect configurations and complex chemical environments. Our findings underscore the significance of both advanced machine learning architectures and comprehensive datasets in advancing uMLIPs. The OMat24 dataset plays a critical role in force predictions, while the eqV2 models demonstrate superior generalization capabilities in defect modeling. Collectively, these advancements position uMLIPs as transformative tools for accelerating materials discovery and design, offering a robust alternative to traditional DFT in computational materials science.

**Acknowledgments**

This work was supported by the Nederlandse Organisatie voor Wetenschappelijk Onderzoek (NWO; the Netherlands Organization for Scientific Research), Domain Science, for access to supercomputing facilities. We also acknowledge the use of the DelftBlue supercomputer provided by the Delft High Performance Computing Center (DHPC; https://www.tudelft.nl/dhpc). We are grateful to Dr. Luca Laurenti for his valuable insights and to Dr. Yury Lysogorskiy for his suggestions on evaluating the computational cost of the various uMLIP models.

**CRediT authorship contribution statement**

**F. S.**: Writing – original draft, Writing – review & editing, Validation, Methodology, Data curation, Conceptualization. **Z. W**: Writing – review & editing, Data curation and Analysis. **K. L.**: Writing – review & editing, Data curation and Analysis. **W. G.**: Writing – review & editing, Data curation and Analysis. **P. D.**: Writing – review & editing, Supervision, Methodology, Funding acquisition, Conceptualization.

**Conflict of Interest**

The authors declare that they have no known competing financial interests or personal relationships that could have appeared to influence the work reported in this paper.

**Data availability**

All uMLIP models are available in their respective GitHub repositories. The DFT datasets generated in this study are not publicly accessible, as they are reserved for ongoing projects. The RANDSPG dataset for Mg can be found from https://doi.org/10.17617/3.A3MB7Z.